\documentclass{ws-procs975x65}

\begin{document}

\title{Binomial multiplicative model of critical fragmentation}

\author{H. Katsuragi, D. Sugino, and H. Honjo}

\address{Department of Applied Science for Electronics and Materials \\
Interdisciplinary Graduate School of Engineering Sciences \\
Kyushu University, 6-1 Kasugakoen, Kasuga, Fukuoka 816-8580, Japan \\
E-mail: katsurag@asem.kyushu-u.ac.jp}



\maketitle

\abstracts{
We report the binomial multiplicative model for low impact energy fragmentation. Impact fragmentation experiments were performed for low impact energy region, and  it was found that the weighted mean mass is scaled by the pseudo control parameter {\it multiplicity}. We revealed that the power of this scaling is a non-integer (fractal) value and has a multi-scaling property. This multi-scaling can be interpreted by a binomial multiplicative (simple biased cascade) model. Although the model cannot explain the power-law of fragment-mass cumulative distribution in fully fragmented states, it can produce the multi-scaling exponents that agree with experimental results well. Other models for fragmentation phenomena were also analyzed and compared with our model.
}


\section{Introduction}
\label{sec:introduction}
Impact fragmentation of brittle solids is a typical nonlinear phenomenon. Small impact cannot make brittle solids cleave. However, large impact produces cracks irreversibly and makes brittle solids fissure to small pieces of fragments. This ubiquitous phenomenon can be seen even in our everyday lives. Thus, many scientists and engineers have studied this issue. As known well, cumulative distribution of fragment mass shows power-law\cite{beysens}. Oddershede et al.\cite{Oddershede} and Meibom and Balslev\cite{Meibom} investigated what controls the exponent of power-law distribution. They found that the exponent is determined by the dimensionality of fractured object. Ishii and Matsushita performed the 1-dimensional fragmentation experiments with long thin glass rods\cite{Ishii}. They dropped the glass rods from various heights. The cumulative distribution obeyed power-law form at high dropping height, and obeyed log-normal form at low one. 

Recently, Kun and Herrmann investigated the damage-fragmentation transition for low impact energy collision by numerical simulation\cite{Kun1}. They used the granular solid disks colliding by a point to each other\cite{Kun2}. The transition from damaged state to fragmentation state was observed by increasing the relative collision speed. They measured the critical exponents of damage-fragmentation transition and realized that scaling-laws of the percolation universality are satisfied near this transition region. On the other hand, {\AA}str\"om et al. studied the low energy fragmentation using the random distorted lattice with elastic beam model and fluid MD model with LJ pair potential\cite{Astrom1}. They corrected that the critical behavior for low energy fragmentation differs from that of percolation. Oddershede et al. said the fragmentation process is a kind of self-organized critical phenomenon\cite{Oddershede}. However, most of experiments examined only on high imparted energy fragmentation. There are no experiments on critical behavior of fragmentation by low inparted energy. 

In order to study the critical fragmentation, we performed simple experiments of fragmentation. We considered a simple binomial multiplicative scenario of critical fragmentation. In this paper, we report on results of detailed analyses on the model. In the next section, experimental results are presented. In Sec.\ \ref{sec:model}, we introduce a binomial multiplicative model and analyze it. In Sec.\ \ref{sec:discussion}, the results are compared with other possible models.

\section{Experiment}
\label{sec:experiment}
We used glass tube samples as 2-D fractured objects. The tube was put between a stainless stage and a stainless plate. A brass weight was dropped to the stainless plate. The falling height was controlled on slightly higher than the point at which samples did not fracture. After fragmentation, we collected fragments and  measured the mass of each fragment with an electronic balance. Fractured tubes have the approximate 2-D geometry ($50$ mm outside diameter, $2$ mm thick, and $50,100,150$ mm length). More detailed experimental setups are described in Ref.\ \ref{ref:katsuragi}. 

\begin{figure}
\begin{center}
\scalebox{1.0}[1.0]{\includegraphics{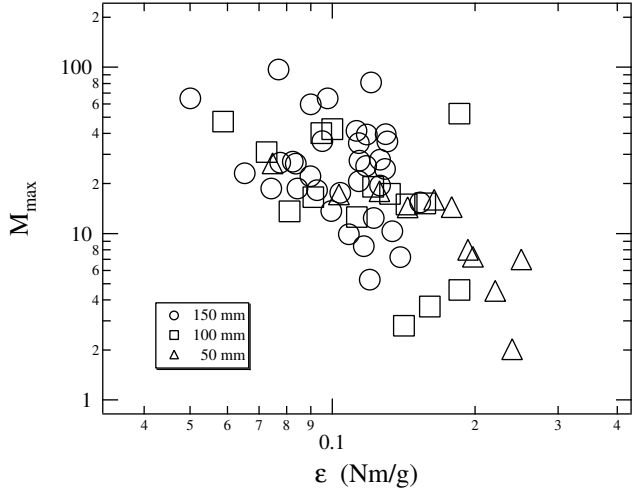}}
\caption{The maximum fragment mass $M_{\max}$ vs. the imparted energy per unit sample mass $\epsilon$. Although the rough correlation between $M_{\max}$ and $\epsilon$ can be seen , it is too fluctuating to discuss quantitatively.}
\label{fig:MmaxVsE}
\end{center}
\end{figure}

According to Kun and Herrmann's result, the control parameter should be the imparted energy per unit sample mass $\epsilon$, and the order parameter should be the maximum fragment mass $M_{\max}$ \cite{Kun1}. The $\epsilon$ is calculated as $\epsilon = M_{w}gh / M_{ob}$, where $M_{w}$, $g$, $h$, and $M_{ob}$ correspond to the mass of falling weight, the gravitational acceleration, the height of falling weight, and the mass of target sample, respectively. The log-log plot of maximum fragment mass $M_{\max}$ vs. imparted energy per unit sample mass $\epsilon$ is shown in Fig.\ \ref{fig:MmaxVsE}. As can be seen in Fig.\ \ref{fig:MmaxVsE}, $M_{\max}$ and $\epsilon$ relate with negative correlation, qualitatively. However, since the data in Fig.\ \ref{fig:MmaxVsE} contain large uncertainties, we cannot discuss quantitatively on the critical scaling by this plot. Therefore, we have to use another parameter to analyze quantitatively.

\begin{figure}
\begin{center}
\scalebox{1.0}[1.0]{\includegraphics{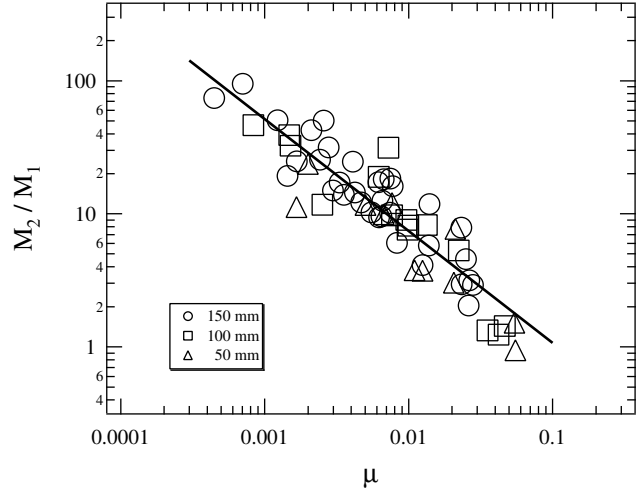}}
\caption{The weighted mean fragment mass $M_2 / M_1$ vs.\ the multiplicity $\mu$. The solid line indicates the form of the power-law fitting $M_2 / M_1 \sim \mu^{-\sigma_1}$ ($\sigma _1= 0.84 \pm 0.05$). Three different size results are fitted by the unified scaling independently of size.}
\label{fig:M2M1}
\end{center}
\end{figure}

\begin{figure}
\begin{center}
\scalebox{1.0}[1.0]{\includegraphics{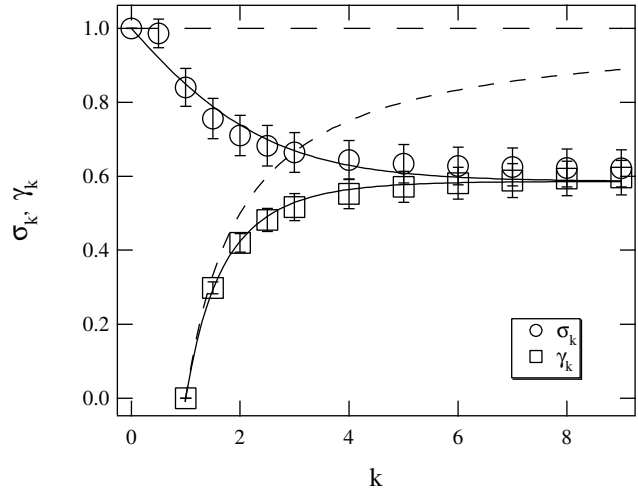}}
\caption{Multi-scaling exponent $\sigma_k$ and $\gamma_k$ obtained from $k$-th order weighted mean fragment mass. The broken lines indicate the case of trivial integral scaling corresponding to $\sigma_k=1$ and $\gamma_k=(k-1)/k$. The solid lines depict results of the binomial multiplicative model at $a=2/3$. The $\sigma_k$ and $\gamma_k$ asymptotically approach to the same value, in large $k$ range. From the definition of Eqs.\ (\ref{eq:sigmak}) and (\ref{eq:gammak}), the values $\sigma_{k=0}$ and $\gamma_{k=1}$ exactly show $1$ and $0$, respectively.}
\label{fig:sigmak}
\end{center}
\end{figure}

Campi proposed a pseudo control parameter {\it multiplicity} $\mu$ in Ref.\ \ref{ref:Campi1}. The $\mu$ is defined as, 
\begin{equation}
\mu=m_{\min}\frac{M_0}{M_1}.
\label{eq:mudef}
\end{equation}
Where $m_{\min}$, $M_0$, and $M_1$ correspond to the smallest limit of fragment mass (we fix it at $0.01$ g), the total number of fragments, and the total mass of the all fragments, respectively. The fragmentation critical point corresponds to the value $\mu=0$ by this definition. In general, we can introduce the $k$-th order moment of fragment mass distribution $M_k$ as, 
\begin{equation}
M_k = \sum_{m} m^{k} n(m),
\label{eq:Mkdef}
\end{equation}
where $n(m)$ is the number of fragments of mass $m$. Certainly, $M_0$ and $M_1$ in Eq.\ (\ref{eq:mudef}) are specific cases of $M_k$ ($k=0$ and $1$, respectively). We consider the $k$-th order weighted mean fragment mass $M_{k+1}/M_{k}$, and assume the scaling,
\begin{equation}
\frac{M_{k+1}}{M_k} \sim \mu ^{-\sigma_k}.
\label{eq:sigmak}
\end{equation}
In Fig.\ \ref{fig:M2M1}, we show the log-log plot of $M_2 / M_1$ as a function of $\mu$. Contrary to the Fig.\ \ref{fig:MmaxVsE}, the data in Fig.\ \ref{fig:M2M1} seem to be fitted by a unified scaling line. The scaling result for $k=1$ is presented as a solid line in Fig.\ \ref{fig:M2M1}. We obtained the nontrivial scaling exponent $\sigma_1=0.84 \pm 0.05$. For other $k$ regime, multi-scaling exponent values of $\sigma_k$ were obtained as shown in Fig.\ \ref{fig:sigmak} (circle marks). In spite of the trivial value $\sigma_{k=0}=1$, $\sigma_k$ varies with $k$, and seems to approach to the nontrivial value ($\simeq 0.6$). 

From the definition of $\gamma_k$ as 
\begin{equation}
\frac{M_k}{M_1} \sim \mu^{-k\gamma_k},
\label{eq:gammak}
\end{equation}
the obtained $\gamma_k$ values are plotted as square marks in Fig.\ \ref{fig:sigmak}. It seems that $\gamma_k$ approaches to the value around $0.6$ again. Of course, Eqs.\ (\ref{eq:sigmak}) and (\ref{eq:gammak}) relate to each other. The relation $\sum^{k-1}_{i=1}\sigma_i=k\gamma_k$ holds for any $k$. Thus, when the $\sigma_k$ has a trivial value $1$ for all $k$, $\gamma_k$ varies as $(k-1)/k$. The trivial curves are shown as broken lines in Fig.\ \ref{fig:sigmak}. In addition, the relation $(k+1)\gamma_{k+1}-k\gamma_k=\sigma_k$ can be computed from Eqs.\ (\ref{eq:sigmak}) and (\ref{eq:gammak}). If the difference between $\gamma_{k+1}$ and $\gamma_k$ becomes small, $\gamma_k$ and $\sigma_k$ approximately have the same value, as seen in Fig.\ \ref{fig:sigmak} for large $k$.  

On the other hand, when the imparted energy was sufficient to fully shatter, many fragments were created and the cumulative distribution of fragment mass obeyed power-law form. Our results suggest the power is $0.5$ about the 2-D fragmentation with the flat impact\cite{Katsuragi}. In this regime, cumulative distribution functions are collapsed by the scaling function written as $N(m)/M_0 \sim f(m/\mu^{-\sigma})$. The function $f(x)$ consists of the scaling part $f(x) \sim x^{-0.5}$and the decaying part due to the finite size effect. The value $0.5$ concurs to the results of Hayakawa\cite{Hayakawa} and {\AA}str\"om et al.\cite{Astrom1}. In contrast, this value is not consistent with the percolation scaling ansatz \cite{Stauffer1}. In the percolation scaling ansatz, the scaling exponent of cluster size cumulative distribution must be greater than $1$. Therefore, we can consider that the universality classes of critical fragmentation and percolation criticality are different each other.

\section{Model}
\label{sec:model}
In order to explain this multi-scaling property, we introduce a simple biased cascade model. A binomial multiplicative process is considered with a unit mass initial condition. Here we consider a asymmetrical cleaving presented by a parameter $a$. We can limit the range of the parameter $a$ as $1/2 \le a \le 1$ by the symmetry of the model. The initial unit mass is divided into two fragments whose masses are $a$ and $1-a$ at first step. This biased cleaving continues some steps until the imparted energy dissipates. In this model, we can easily calculate the exponents $\sigma_k$ and $\gamma_k$ from Eqs.\ (\ref{eq:sigmak}) and (\ref{eq:gammak}) as 
{\setcounter{enumi}{\value{equation}}
\addtocounter{enumi}{1}
\setcounter{equation}{0}
\renewcommand{\theequation}{\theenumi\alph{equation}}
\begin{equation}
\frac{a^{k+1} + (1-a)^{k+1}}{a^k + (1-a)^k} = 2^{-\sigma_k}, 
\label{eq:modelsigmak} 
\end{equation}
\begin{equation}
a^k + (1-a)^k = 2^{-k\gamma_k},
\label{eq:modelgammak}
\end{equation}
\setcounter{equation}{\value{enumi}}}
or more explicit forms as,  
{\setcounter{enumi}{\value{equation}}
\addtocounter{enumi}{1}
\setcounter{equation}{0}
\renewcommand{\theequation}{\theenumi\alph{equation}}
\begin{equation}
\sigma_k = -\frac{\ln [a^{k+1}+(1-a)^{k+1}] - \ln [a^k + (1-a)^k]}{\ln 2}, \label{eq:binomialsk} 
\end{equation}
\begin{equation}
\gamma_k = -\frac{\ln [a^k + (1-a)^k]}{k\ln 2}. \label{eq:binomialgk}
\label{eq:modelgammak}
\end{equation}
\setcounter{equation}{\value{enumi}}}
If we choose a value $a=2/3$, the $\sigma_k$ and $\gamma_k$ become the values depicted by the solid lines in Fig.\ \ref{fig:sigmak}. One can confirm the pretty good agreement with experimental data. The trivial case presented by broken lines in Fig.\ \ref{fig:sigmak} corresponds to the case $a=1/2$. In this case, all fragments at each step are perfectly equal. In the case $a \neq 1/2$, the fragment size distribution has variation and exhibits multifractal scaling.

This model is so simple that we can calculate the fragment mass and the number of fragments exactly. We consider the $s$-th step, and suppose the fragments in which the factor $a$ (or $1-a$) works $t$ (or $s-t$) times. In such fragments, the mass $m_s(t)$ is written as, 
\begin{equation}
m_s(t)=a^t(1-a)^{s-t}, \;\;\;\; (\frac{1}{2} \leq a \leq 1).
\label{eq:m}
\end{equation}
And the number of fragments $n_s(t)$ is described as, 
\begin{equation}
n_s(t)=\frac{s!}{t!(s-t)!}.
\label{eq:n}
\end{equation}

\begin{figure}
\begin{center}
\scalebox{1.0}[1.0]{\includegraphics{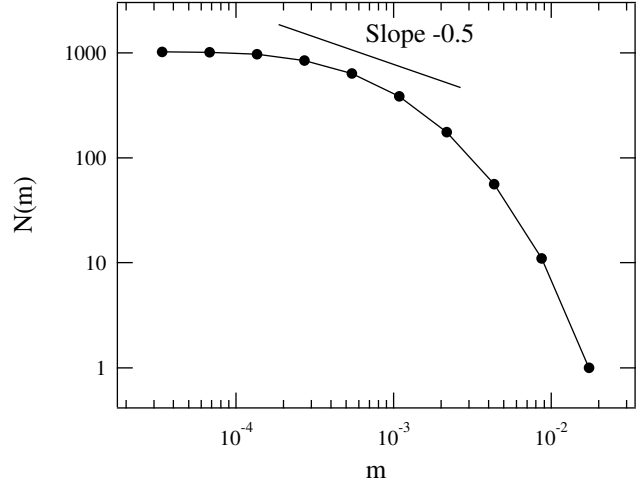}}
\caption{The cumulative fragment mass distribution for the binomial multiplicative model with $a=2/3$ and $s=10$.}
\label{fig:Nm}
\end{center}
\end{figure}

Since we are interested in the cumulative distribution of fragment mass, the cumulative number of fragments $N_s(t)$ is calculated as 
\begin{equation}
N_s(t)=\int_t^{\infty} n_s(t')dt'=\sum_{i=t}^{s}\frac{s!}{i!(s-i)!}.
\label{eq:Nm}
\end{equation}
In Fig.\ \ref{fig:Nm}, we show the cumulative distribution of fragment mass obtained from the model. The parameters are taken as $a=2/3$ and $s=10$. The line of slope $-0.5$ corresponding to the experimental result is also shown as a solid line in Fig.\ \ref{fig:Nm}. Unfortunately, clear power-law, which follows the experimental data, cannot be observed. However, the distribution curve in Fig.\ \ref{fig:Nm} seems to include the region of slope $-0.5$. 
\begin{figure}
\begin{center}
\scalebox{1.0}[1.0]{\includegraphics{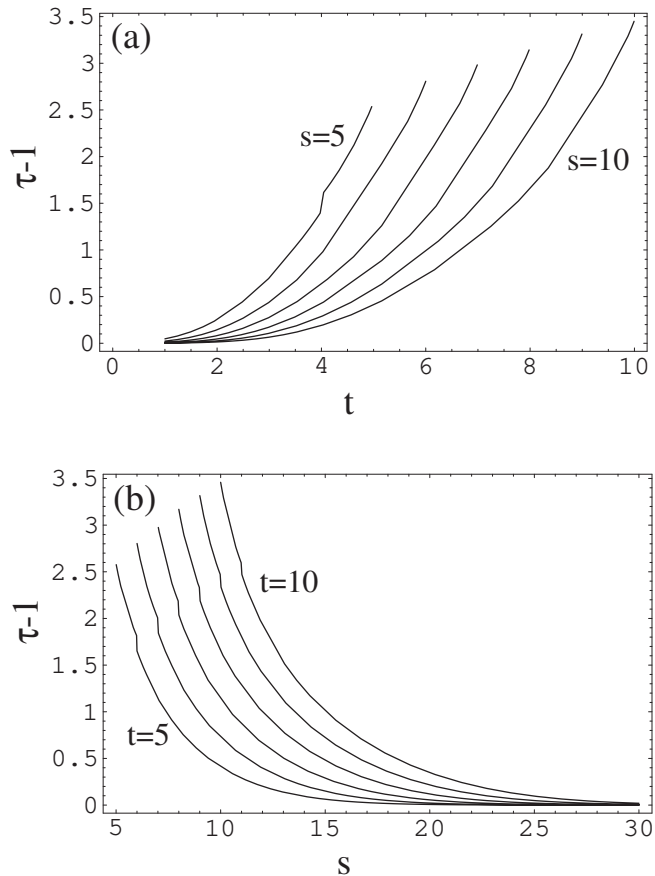}}
\caption{The local scaling exponent of cumulative fragment mass distribution $\tau-1$. (a) The relations between $\tau-1$ and $t$ are shown. Each curve corresponds to the case $s=5,6, \cdots, 10$ from left to right. (b) The relations between $\tau-1$ and $s$ are shown. Each curve corresponds to the case $t=5,6, \cdots, 10$ from left to right. }
\label{fig:tau}
\end{center}
\end{figure}
In this model, we can calculate the local power $\tau-1$ directly by the relation,
\begin{equation}
\frac{N_s(t-1)}{N_s(t)}= \left[ \frac{m_s(t-1)}{m_s(t)} \right] ^{-(\tau-1)}.
\label{eq:t1}
\end{equation}
Solving the Eq.\ (\ref{eq:t1}), we obtain the exact form of $\tau-1$ as follows, 
\begin{equation}
\tau-1=-\frac{\ln \left[ {\displaystyle \frac{N_s(t-1)}{N_s(t)}} \right] }{\ln \left[ {\displaystyle \frac{m_s(t-1)}{m_s(t)}} \right] }=-\frac{\ln \left[ \frac{{\textstyle \sum_{i=t-1}^{s}\frac{s!}{i!(s-i)!}}}{{\textstyle\sum_{i=t}^{s}\frac{s!}{i!(s-i)!}}} \right]}{\ln \left[ {\displaystyle \frac{a^{t-1}(1-a)^{s-t+1}}{a^t(1-a)^{s-t}}} \right] }.
\label{eq:t2}
\end{equation}
We show the relations among $\tau-1$, $s$, and $t$ in Fig.\ \ref{fig:tau}. As can be seen in Fig.\ \ref{fig:tau}, the lower limit of the local slope $\tau-1$ is $0$, and it has a divergent tendency. The value $0.5$ is not a particular one.

\section{Discussion}
\label{sec:discussion}
In Sec.\ \ref{sec:experiment}, we concluded that the universality of the critical fragmentation differs from that of percolation. Instead, the weighted mean fragment mass was studied to reveal the universality of the critical fragmentation. It indicates the multi-scaling nature and is modeled by the simple binomial multiplicative model. There are some other candidates for the critical fragmentation. From now on, we discuss and compare them. 

Similar multiplicative model for turbulent flows was proposed by Meneveau and Sreenivasan\cite{Meneveau1}. They investigated the energy cascade of eddies, and obtained good agreement with experimental data at $a=0.3$ (in their paper, the corresponding parameter was written as $p_1$). This value slightly coincides to ours $1-a=1/3$. The same physical mechanism might dominate both cascades of fragmentation and turbulence. 

\begin{figure}
\begin{center}
\scalebox{1.0}[1.0]{\includegraphics{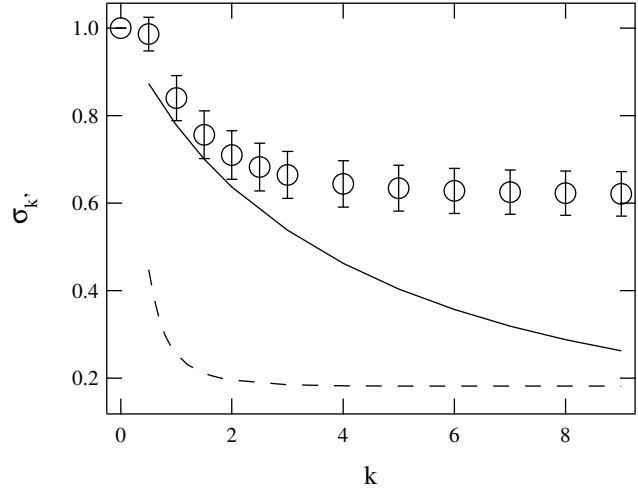}}
\caption{Comparison between the experimental data and the other considerable models. The experimental data are shown as circle marks. The exponent $\sigma_k$ obtained from the symmetric (isosceles triangle) distribution model is presented by the solid line. The broken line indicates the result of the modified partial remaining model at $s=10$.}
\label{fig:sigmakd}
\end{center}
\end{figure}

We can fit the data by $a=2/3$ very well indeed, however, the reason of  symmetry breaking by $a \neq 1/2$ is not understood well. While the model always requires the exact asymmetry presented by $a$, the cleaving point might distribute. We can consider the simple distributed model as described below. We set the unit mass segment initial condition again, and consider the probability $p(x)dx$ which presents the cleaving point in the range from $x$ to $x+dx$ at each step. We assume the symmetrical distribution as $p(x)=4x\;(0 \leq x \leq1/2)$, and $4-4x \; (1/2 \leq x \leq 1)$. This is one of the simplest distribution presented by isosceles triangles. The normalization condition of this model is $\int_0^1p(x)dx=1$. In this model, we can calculate the expectation value of the cleaving point $x$ (or $1-x$)  as,
\begin{equation}
\int_0^{\frac{1}{2}} (1-x) 4x dx + \int_{\frac{1}{2}}^1 x (4-4x) dx = \frac{2}{3}.
\label{eq:pxdx}
\end{equation}
Note that we cannot distinguish the cleaving state ($x, 1-x$) and the state($1-x, x$). Thus, the $x$ can be limited in the region $1/2\leq x \leq 1$. The expectation value is nearly the same as one ($a=2/3$) of the above mentioned multiplicative model. We can also calculate the $k$-th order moment $M_k$ as, 
\begin{equation}
M_k = \int_0^1 [x^k + (1-x)^k]p(x) dx  = \frac{8}{(k+1)(k+2)} \left[ 1 - \left(\frac{1}{2} \right)^{k+1} \right].
\label{eq:Mkdist}
\end{equation}
We show the $\sigma_k$ computed from the Eqs.\ (\ref{eq:Mkdist}), (\ref{eq:sigmak}), and (\ref{eq:mudef}) as a solid line in Fig.\ \ref{fig:sigmakd}. The result does not supply the agreement with the experimental data, particularly in large $k$ range. 

Matsuhita\cite{Matsushita1} and Turcotte\cite{Turcotte1} introduced the model for power-law fragmentation. Matsushita examined the model in which each fragment cleaves into 4 pieces at each step. Then 1 piece does not  break any more, and  the other 3 pieces cleave into 4 sub-pieces at next step. The same procedure works upon all sub-pieces at each step. According to his model, the exponent of the power-law cumulative distribution of fragment mass becomes $\tau-1 =\ln 3/\ln 4 \simeq 0.79$. We can easily modify this model to the case $\tau-1 = \ln 2/\ln 4=1/2$ by changing the remaining piece number 1 into 2. In this condition, we can calculate the $M_k$ for this modified version of the partial remaining model at $s$-th step as,
\begin{equation}
M_k = \sum_{i=1}^{s} 2^i \left( \frac{1}{4} \right) ^{ik}.
\label{remainMk}
\end{equation}
Then, the $\sigma_k$ can be computed again, however, the value of $\sigma_k$ depends not only on $k$, but also on $s$. We show the $\sigma_k$ obtained from this model at $s=10$ as a broken line in Fig.\ \ref{fig:sigmakd}. This model also cannot provide the appropriate curve of the $\sigma_k$. Thus the exact $a=2/3$ binomial multiplicative model seems to be the most possible model in terms of multi-scaling exponents $\sigma_k$.

\section{Conclusions}
\label{sec:conclusion}
We investigated the criticality of brittle fragmentation. Some models to interpret the experimental result are proposed. The exact binomial multiplicative model can produce the adequate approximation for the exponent $\sigma_k$. And the cumulative distribution obtained from the model is not so worse. However, it requires that the cleaving point is exactly at $a$. Since the isosceles triangle model has non-divergent standard deviation, the distribution of fragment mass resulting from the model must approach to the log-normal distribution due to the central limit theorem\cite{Matsushita2,Diehl1}. In addition, its $\sigma_k$ differs from the experimental data, particularly in large $k$ region. The modified partial remaining model can explain the experimental value of the exponent $\tau$ very well. However, the $\sigma_k$ from the model shows large discrepancy from the experimental data. Each model has merits and demerits. The totally sufficient model is not presented yet. Furthermore, these scaling exponents will depend on detail load condition and dimensionality of fractured object. More detailed experiments and analyses are necessary to solve the criticality of brittle fragmentations completely.

\section*{Acknowledgments}
The authors would like to thank Professor J. Timonen , Dr.\ J. A. {\AA}str\"om, Professor S. Ohta, Professor H. Sakaguchi, and Professor K. Nomura for useful discussion and comments.

\end{document}